\documentclass[aps,prl,twocolumn,superscriptaddress]{revtex4-1}
\usepackage{graphicx}
\usepackage{amsmath}
\usepackage{xcolor}
\usepackage{physics}
\usepackage{hyperref}
\usepackage[all]{hypcap}
\hypersetup{backref,
	pdfpagemode=FullScreen,
	colorlinks=true, citecolor=blue}
\usepackage[sort&compress]{natbib}

\usepackage{color}
\usepackage{comment}
\newcommand{\C}{\color{magenta}}

\renewcommand{\C}[1]{}

\begin{document}
	\title{Probing statistics of coherent states by continuous wave mixing on a single artificial atom}	
	\author{A. Yu. Dmitriev}
	\email[]{dmitrmipt@gmail.com}
	\affiliation{Moscow Institute of Physics and Technology, 141700 Dolgoprudny, Russia}
	
	\author{R. Shaikhaidarov}
	\affiliation{Physics Department, Royal Holloway, University of London, Egham, Surrey TW20 0EX, United Kingdom}
	
	\author{T. H\"{o}nigl-Decrinis}
	\affiliation{Physics Department, Royal Holloway, University of London, Egham, Surrey TW20 0EX, United Kingdom}
	\affiliation{National Physical Laboratory, Teddington, TW11 0LW, United Kingdom}
	
	\author{S. E. de Graaf}
	\affiliation{National Physical Laboratory, Teddington, TW11 0LW, United Kingdom}
	
	\author{V. N. Antonov}
	\affiliation{Physics Department, Royal Holloway, University of London, Egham, Surrey TW20 0EX, United Kingdom}
	\affiliation{Skolkovo Institute of Science and Technology, Nobel str. 3, 143026 Moscow, Russia}
	\affiliation{Moscow Institute of Physics and Technology, 141700 Dolgoprudny, Russia}
	
	\author{O. V. Astafiev}
	\email[]{Oleg.Astafiev@rhul.ac.uk}
	\affiliation{Physics Department, Royal Holloway, University of London, Egham, Surrey TW20 0EX, United Kingdom}
	\affiliation{National Physical Laboratory, Teddington, TW11 0LW, United Kingdom}
	\affiliation{Moscow Institute of Physics and Technology, 141700 Dolgoprudny, Russia}

	\begin{abstract}
We study four- and higher-order wave mixing of continuous coherent waves on a single superconducting artificial atom. Narrow side peaks of different orders of nonlinearity resulting from elastic multi-photon scattering on the atom are observed and investigated. We derive an analytical expression for the peak amplitudes and show that the ratio of any two adjacent peaks is a function of driving amplitudes and detuning. 
This is attributed to the photon distribution in the coherent states and provides a measure for characterisation of photon statistics in non-classical coherent waves. We also demonstrate an Autler-Townes-like splitting of side peaks, the magnitude of which scales with the scattering order. 

		\end{abstract}
	
	\date{\today}
	
	\maketitle
{\C{Introduction.}} The field of superconducting quantum circuits \cite{clarke2008superconducting,Devoret_SC_Outlook} strongly coupled to either confined \cite{wallraff2004strong} or propagating \cite{Astafiev2010resonance} electromagnetic waves is rapidly developing area of experimental physics. It turned out to be an especially advantageous toolkit for the demonstration of quantum optical phenomena in the microwave frequency domain \cite{Microwave_optics_review}, offering conditions which are not acheivable with natural atoms. Some striking examples are the demonstration of strong \cite{Abdumalikov_EIT,Delsing2013microwave1Dspace,Astafiev-quantum-amplifier} and ultra-strong \cite{niemczykUSC, FornDiaz_USC,Lupasku2017ultrastrong} coupling of light to a single qubit, the coupling of qubits by virtual photons \cite{vanLoo1494}, the dynamical Casimir effect \cite{wilson2011observation}, sources of single photons \cite{peng2016tuneable,Shaped_SPS} and entangled propagating photons \cite{Wallraff_entangledPhotons}, lasing \cite{astafiev2007single} and amplification \cite{Astafiev-quantum-amplifier} with a single artificial atom.  A range of specific effects related to the intrinsic nonlinearity of an atom were demonstrated, for example, the phase shift acquired by single propagating photons \cite{Delsing-giant-Kross-Kerr}, or the preparation of cat states in the cavity \cite{kirchmair2013observation}. In general, the nonlinearity is a tool for the implementations of quantum gates, as described in various proposals \cite{Fredkin, Brod_CPHASE, munro2005weak, WeakNon}. Therefore, the nonlinear regimes of light-matter interactions in circuit-QED are of specific importance and interest for both fundamental quantum optics and quantum information processing.

{\C{What is wave mixing.}} 
Wave mixing in optical media is one of the basic nonlinear parametric processes. In particular, four-wave mixing occurs due to the third-order susceptibility $\chi^{(3)}$ of a media, which gives the polarization term  $P^{(3)} = \chi^{(3)}E^3$, proportional to third power of electric field $E$. If three electromagnetic waves with frequencies $\omega_1$, $\omega_{2}$, $\omega_{3}$ propagate through the media, a number of additional waves appear with frequencies $\pm\omega_i\pm\omega_j\pm\omega_k$, where $i,j,k=\{1,2,3\}$ \cite{boyd2003nonlinear}, and their amplitudes and phases depend on the amplitudes and phases of the initial waves. This also means that three photons are required to allow corresponding scattering processes. With higher $2p+1$ odd-order nonlinearity, where $p$ is a non-negative integer number, the processes is characterised by $\chi^{(2p+1)}$ and one extra photon is generated as a result of interaction between $2p+1$ photons. A wide range of applications of wave mixing in various systems includes generation of squeezed states of light \cite{slusher1985observation, Eichler_squeez,Toyli2016ResSqueez}, parametric amplification \cite{castellanos2008amplification}, frequency conversion and generation of frequency combs \cite{kippenberg2011microresonator}. For superconducting quantum circuits, the effect of wave mixing \cite{dmitriev2017quantum} of short microwave pulses was observed and characterised. Very recently, it was shown that a strongly driven two-level system could amplify a weak probe in between the components of the Mollow triplet due to four-photon processes \cite{wen2017reflective}. Three-wave mixing was also theoretically described for cyclic artificial atoms \cite{liu2014controllable}, and experimentally observed \cite{Hoenigl2017} on a single three-level circuit.

\begin{figure}[h]
	\includegraphics[width=1\columnwidth]{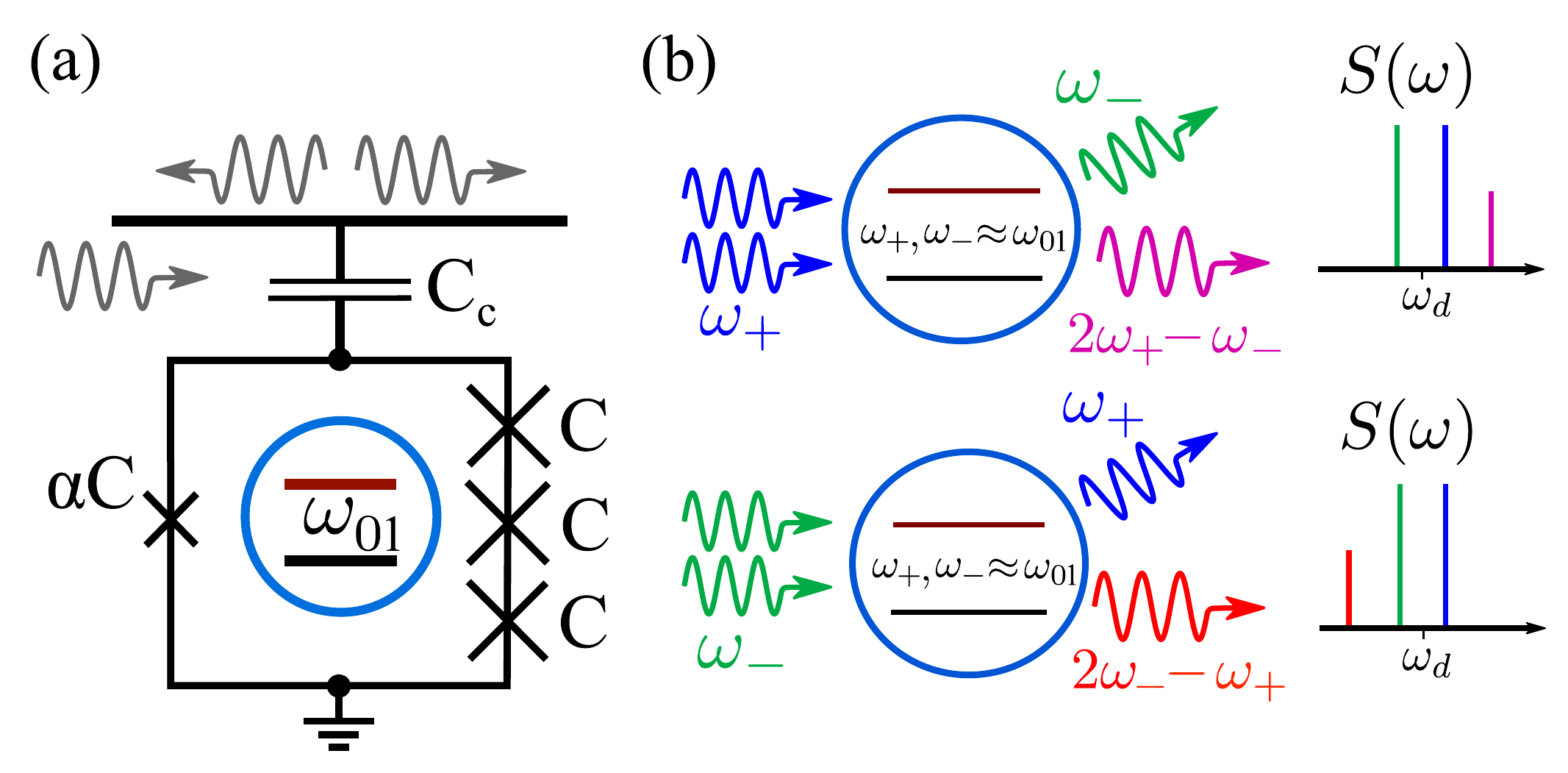}
	\caption{\textbf{(a)} Schematics of the device. The artificial atom is a 4-junction flux qubit with three nearly identical junctions with capacitance $C$ and one junction with capacitance $\alpha C$, where $\alpha = 0.43$. The qubit is coupled to a coplanar waveguide with capacitance $C_c$. \textbf{(b)} The wave mixing of two tones on a single two-level system. With the two-photon absorption and one photon emission at frequencies $\omega_\pm$ and $\omega_\mp$ one more photon is generated at frequency $2\omega_\pm - \omega_\mp$ resulting in corresponding side frequency emissions. }
	\label{Intro_WM}
\end{figure}

{\C{What we have done.}}
In our work, we study the wave mixing of continuous coherent waves on a single superconducting qubit as a two-level artificial atom strongly coupled to a coplanar waveguide. An atom is irradiated by two propagating microwaves at frequencies close to resonance. We measure narrow sideband spectral peaks \cite{Mollow} attributed to elastic multi-photon scattering processes of four-, six- and higher-orders. With two waves any order process is allowed due to the finite probability of finding any photon number in the coherent states. The side peak intensities depend on the order of nonlinear processes, incident wave amplitudes, and their detuning from the atomic transition. An interesting feature is that with the coherent wave scattering the ratio between consequent peak intensities is independ of the peak orders. 
Another interesting finding is that with a strong drive, the side peak intensities exhibit splitting in frequency domain with the magnitude proportional to the driving amplitudes. This is similar to the Autler-Townes splitting, (studied also in superconducting circuits \cite{Baur,ATS_3LS,Abdumalikov_EIT,PengATS}), however, the corresponding splitting magnitude is inversely proportional to the order of the scattering process. We derive an analytical expression for the spectra and obtain a good agreement with the experimental data.

{\C{Theoretical background.}}
We begin with considering a two-level atom with transition frequency $\omega_{01}$, see Fig.~\ref{Intro_WM}(a). The atom is strongly coupled to a transmission line (open space) with the radiative relaxation rate $\Gamma_1$ due to the photon emission into the line. The strong coupling condition implies that non-radiative relaxation (without emission of the photon to the line) $\Gamma_1^{nr}$ and pure dephasing rates $\gamma$ are smaller than $\Gamma_1$. A driving monochromatic wave with frequency $\omega_d$ and wavevector $k$ described by voltage amplitude $V_0e^{-i\omega_d t+i kx}$ propagates through the waveguide and scatters on the atom located at $x = 0$. As a result, the wave is scattered elastically and inelastically either forward or backward. The amplitude of the elastically scattered wave $V^{sc} e^{-i\omega_d t +ik|x|}$ is expressed as 
\begin{equation}
V^{sc} = -\frac{i\Gamma_1}{\hbar \mu}\langle \sigma^-\rangle, 
\label{Vsc}
\end{equation}
where $\sigma^-$ is the atomic state annihilation operator and $\mu$ is the atomic dipole moment. Here we will not consider an inelastically scattered radiation. By finding the stationary solution of the master equation for the atom with the external drive, it can be shown \cite{Astafiev2010resonance} that the amplitude of the elastically scattered wave $V^{sc} e^{-i\omega_d t +ik|x|}$ is expressed as
\begin{equation}
\label{refl}
V^{sc} = -r V_0 = -\frac{V_0}{2} \frac{\lambda\Gamma_1}{|\lambda|^2+\Omega^2\Gamma_2/\Gamma_1}, 
\end{equation}
where $\lambda = \Gamma_2 + i\Delta\omega$, $\Delta\omega = \omega_d - \omega_{01}$ is detuning, $\Gamma_2 = \frac{\Gamma_1 + \Gamma_1^{nr}}{2} + \gamma$ is the full dephasing rate, $\Omega = \mu V_0/\hbar$ is the driving amplitude of the incident wave, and $r$ is reflection coefficient. One important consequence of Eq.~(\ref{refl}) is that in a week driving limit ($\Omega \ll \Gamma_1$) with $\Gamma_2 = \Gamma_1/2$ (ideal strong coupling: $\Gamma_1^{nr} = \gamma = 0$) and $\lambda = \Gamma_1/2$ at $\Delta\omega = 0$, the scattered wave is equal to the incident wave in amplitude but negative in sign: $V^{sc} = - V_0$.

Next, we generalise the problem to the scattering of two coherent waves at frequencies $\omega_+ = \omega_d + \delta\omega$ and $\omega_- = \omega_d - \delta\omega$, as depicted in Fig.~\ref{Intro_WM}(b), where the frequency shift is small: $\delta\omega \ll \Gamma_1$. 
The mixing processes can be described in terms of multi-photon elastic scattering. In particular, Fig.~\ref{Intro_WM}(b) illustrates four-wave mixing processes ($2p+1 = 3$). The upper panel describes a mechanism of a photon generation at $2\omega_+ - \omega_-$ as a result of the four-photon process: two photons from the $\omega_+$--mode are absorbed and two photons are emitted, one at $\omega_-$ and one at $2\omega_+ - \omega_-$
. The lower panel represents a symmetric process with emission of  a photon at $2\omega_- - \omega_+$. These two processes are called degenerate four-wave mixing and can be spectroscopically detected by observation of side spectral peaks at corresponding frequencies. As a probability to find an arbitrary number of photons in coherent states is finite, any higher-order processes take place. They result in creation of the spectral components at $\omega_{\pm (2p+1)} =  (p+1)\omega_{\pm} - p\omega_{\mp}$ as an outcome of scattering with $2p+1$ photons involved, where $p \geq 0$ is an integer. We are measuring the continuous mixing of two coherent waves on a superconducting quantum system strongly coupled to a transmission line. 

The artificial atom in our device is a flux qubit coupled to a coplanar waveguide by capacitance~$C_c = 2$~fF (Fig. \ref{Intro_WM}(a)), which also effectively shunts the $\alpha$-junction \cite{dmitriev2017quantum}. The  persistent current is estimated to be quite small, $I_p = 52$~nA, however, the anharmonicity $\omega_{12} - \omega_{01} \approx 2\pi\times1.5$~GHz is still large to not account for higher levels. Other important parameters are the energy splitting at the degeneracy point $\Delta_q = 2\pi\times7.30$~GHz and rates of relaxation and decoherence $\Gamma_1/2\pi = 2.2$~MHz and $\Gamma_2/2\pi = 1.1$~MHz measured at $\omega_{01}=\Delta_q$. 
To measure the transmission coefficient of a waveguide with the embedded qubit, we built a standard low-temperature setup described elsewhere \cite{Astafiev-quantum-amplifier}. The qubit is located at the 15 mK-flange in a dilution refrigerator. The input microwave signals with very narrow spectral widths of $\sim1$~Hz are delivered to the chip via coaxial lines with attenuators at different temperature stages used to suppress the room-temperature blackbody radiation. The output signals go through a microwave isolator and are amplified by a cryogenic HEMT amplifier situated at 4K stage of the refrigerator and then by two microwave amplifiers at room temperature. 

\begin{figure}[tbh]
	\includegraphics[width=1\columnwidth]{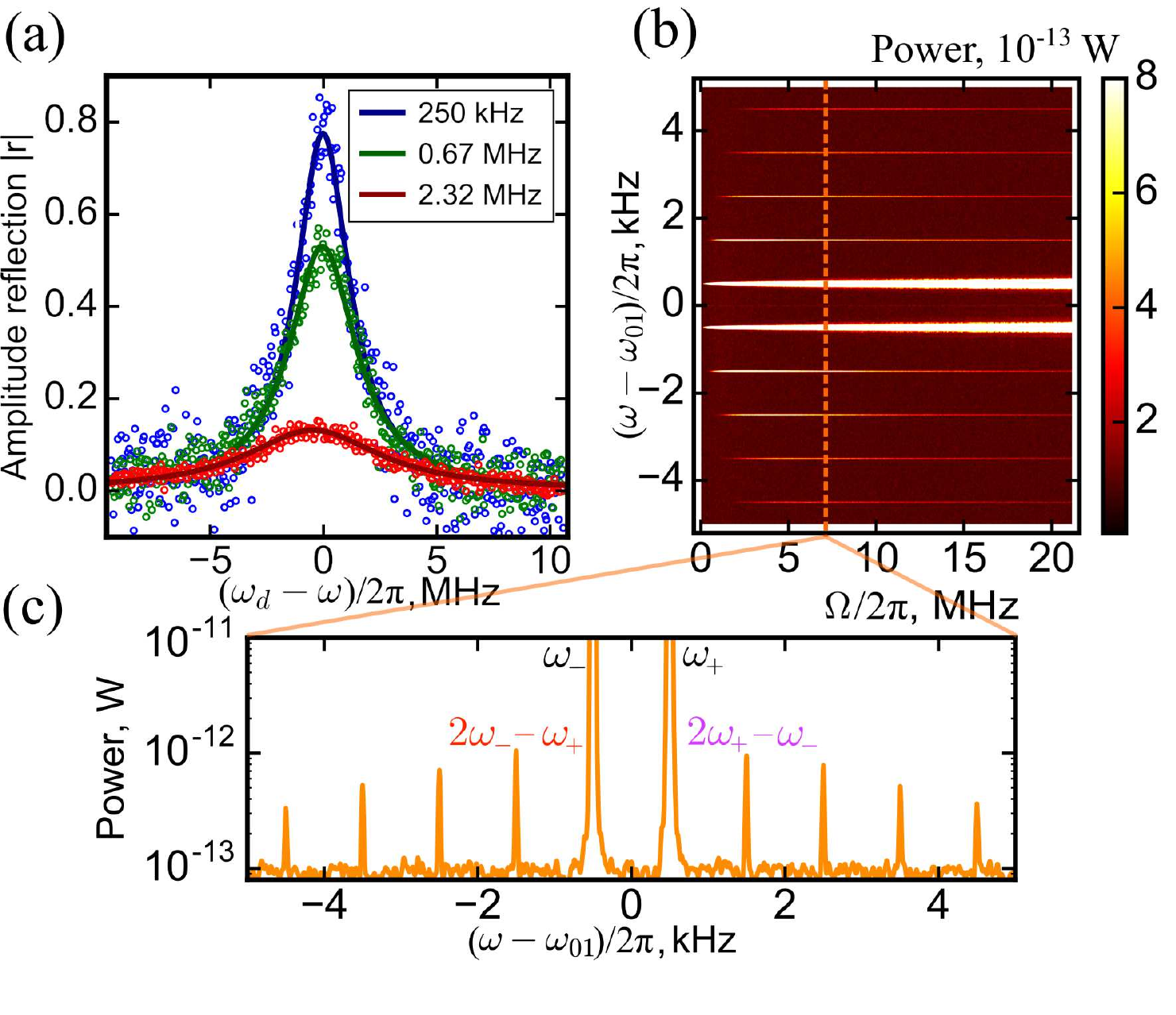}
	\caption{ \textbf{(a)} The  single frequency wave elastically scattered from the artificial atom: blue dots correspond to a weak drive. At higher powers, the atom is saturated and the reflection is reduced (green and red dots). Solid lines are fits made using Eq.~(\ref{refl}). The legend captions are values of $\Omega/2\pi$ extracted from the fit. \textbf{(b)} The spectra of coherently scattered radiation measured by a spectrum analyser when driving tones are resonant with the qubit, plotted as a function of the amplitude of both tones: $\Omega_+ = \Omega_- = \Omega$. \textbf{(c)} An example of a typical spectrum.}
	\label{Omegas}
\end{figure}

The transmitted waves can be measured by a vector network analyzer or by a spectrum analyzer.  Preliminary calibration is performed by measuring the transmission coefficient $t = V_{tr}/V_0$~of a single microwave tone of frequency $\omega_d \approx \omega_0$, and using the relation $r+t = 1$. We calculate the value of the reflection coefficient $r$ and fit it using Eq.~(\ref{refl}). The result is shown in Fig.~\ref{Omegas}(a). By fitting the peak, we obtain a value of $\Omega$ for a certain output level of the microwave generator and thereby we can determine the amplitude of the driving signal, see Fig.~\ref{Omegas}(a). 

\begin{figure}[tbh] 
	\includegraphics[width=1\columnwidth]{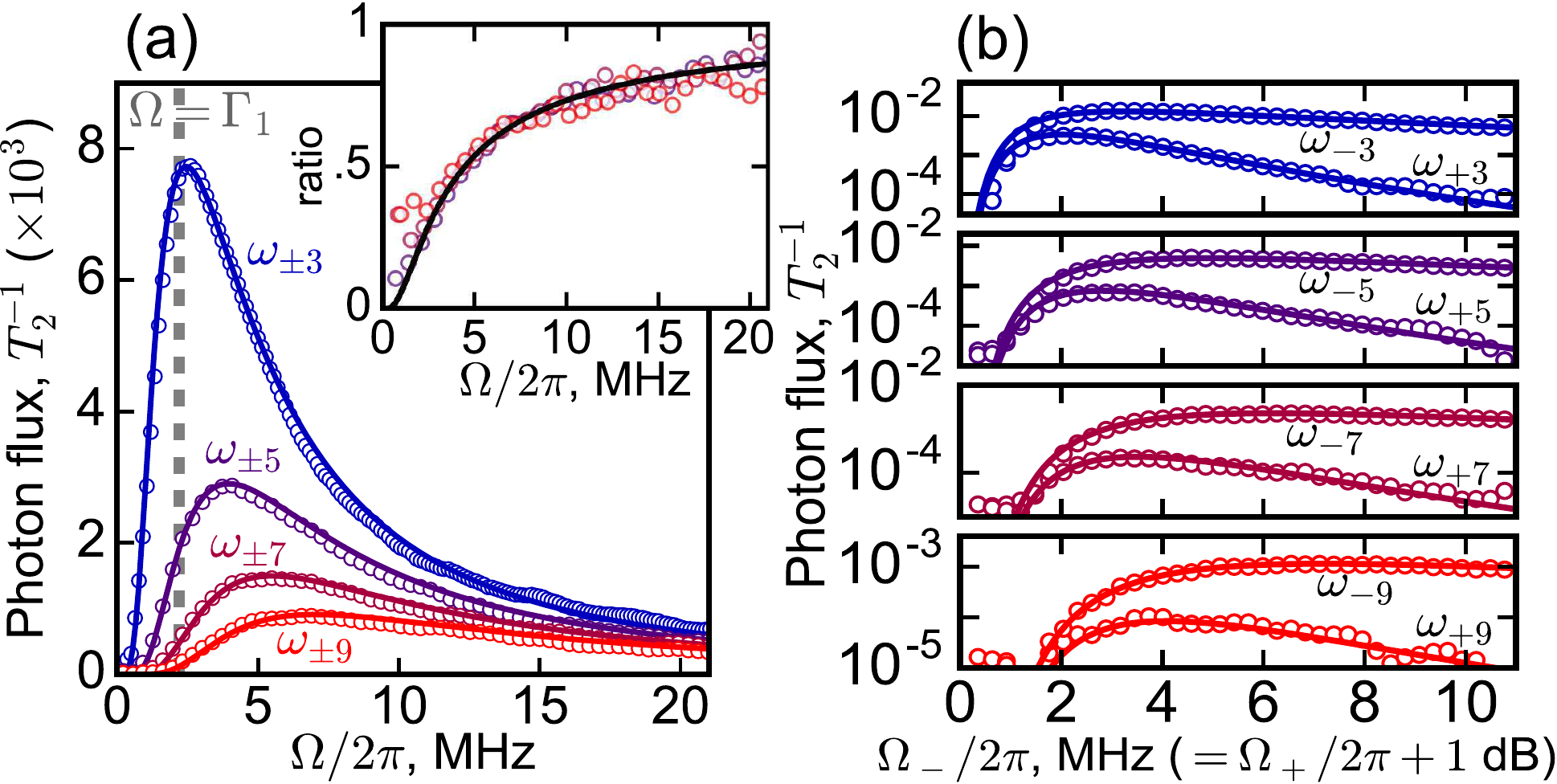}
	\caption{\textbf{(a)} Sideband spectral components of elastically scattered waves. Experimental points obtained with $\delta\omega=5$~kHz, are fitted with Eq.~(\ref{an_eq}) (solid lines) with parameters $\Gamma_1 = 2.2$~MHz, $\Gamma_2=1.1$~MHz, $\Delta\omega=0$, $\Omega_+=\Omega_- =\Omega$ and $p=1,2,3,4$ for each curve, respectively. Inset shows the ratio of photon fluxes in components of consequent order $2p+1$ and $2p+3$ for $p=1,2,3$. The black line represents the direct evaluation of ratios from Eq.(\ref{intensities}). \textbf{(b)} The wave mixing for asymmetric driving signals: $\Omega_-$~is larger by 1 dB than~$\Omega_+ $. The positive side components are a few times higher than the negative ones. This demonstrates the high sensitivity of wave mixing to the disbalance of the driving amplitudes.}
	\label{Peaks_WM_with_fit}
\end{figure}

{\C{Experiment.}} 
For the demonstration of wave mixing, we tune the qubit to the degeneracy point $\omega_{01} = \Delta_q$ and apply two microwaves at frequencies $\omega_+ = \omega_d + \delta\omega$ and $\omega_- = \omega_d - \delta\omega$, as depicted in Fig.~\ref{Omegas}(b). The detuning $\delta\omega$ is typically chosen to be 1 - 100~kHz $\ll \Gamma_1,\Gamma_2$, therefore both tones are within the width of the resonance line ($\sim\Gamma_2$) with a qubit but still easily distinguishable. By measuring the spectrum of the output signal, we observe many side spectral components at frequencies $\omega_{\pm(2p+1)} = \omega_d \pm(2p+1) \delta\omega$, where $p>0$ is an integer, see Fig.~\ref{Omegas}(b, c). 
Figure~\ref{Peaks_WM_with_fit}(a) demonstrates the side peak amplitudes for $1< p < 4$ (up to 9-photon process) with $\Omega_+ = \Omega_-$ as a function of the driving amplitude: left and right hand-side peaks are of equal amplitudes. Depending on he peaks reach maxima and then decay. With increasing the order of the processes the maxima are obtained for a higher driving amplitude. 
We interpret this in the following way. The peak order ($2p+1$) corresponds to the number of interacting photons. The photon absorption/emission rate is determined by the Rabi-oscillation frequency, equivalent to the driving amplitude $\Omega$. The characteristic interaction time determined by the system coherence is $\tau\approx\Gamma_2^{-1}$. Therefore, to efficiently absorb/emit $2p+1$ photons, one needs to drive the system with the amplitude $2\Omega\tau \approx 2p+1$ (here we take $\Omega_- + \Omega_+ = 2\Omega$) and we obtain {$\Omega_{max} \approx \Gamma_1 (2p+1)/4$} (we take $\Gamma_2 = \Gamma_1/2$). 

We also investigate how the intensities of the side peaks depend on the difference in amplitudes of the driving waves, when $\Omega_- \neq \Omega_+$. To illustrate that we vary both driving amplitudes while keeping the amplitude $\Omega_{-}$ 1 dB (1.26 times in amplitude) higher than $\Omega_+$ and measure the side-band components (Fig.~\ref{Peaks_WM_with_fit}(b)). The symmetry is now broken and the intensities at $\omega_{-(2k+1)}$ becomes several times larger than at $\omega_{+(2k+1)}$. The processes generating positive frequency peaks become less probable than the ones resulting in negative frequency components. This is a direct consequence of having more photons in the $\omega_-$--mode. 
\begin{figure*}[tbh]
	\includegraphics[width=1.5\columnwidth]{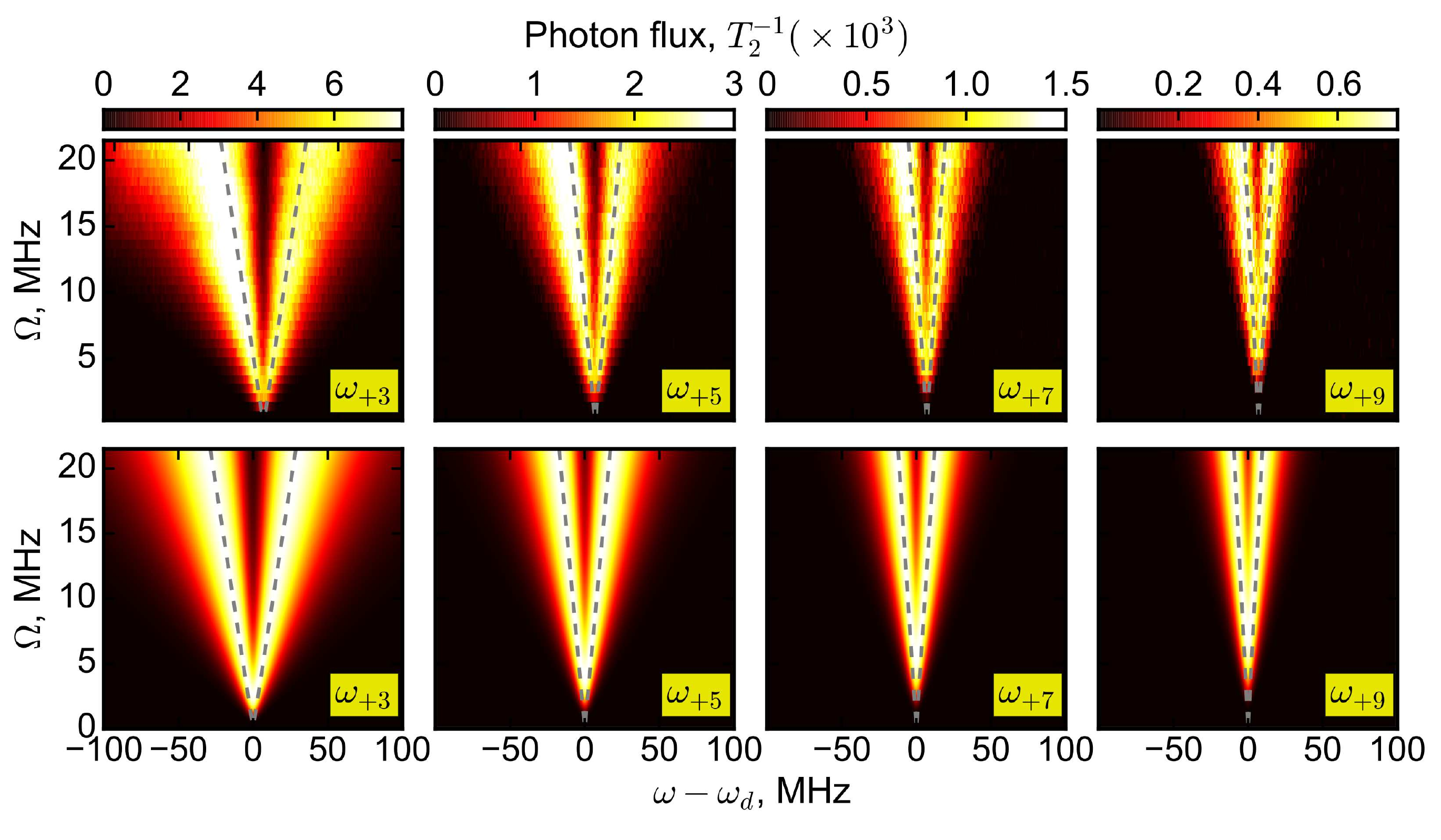}
	\caption{ The Autler-Townes-like splitting of side spectral components of scattered radiation as a function of the central frequency detuning and driving amplitudes $\Omega$. The upper panels represent measured splittings. Lower panels represent analytical calculations with Eq.~(\ref{intensities}). Grey dashed lines are the guidelines corresponding to $\Delta\omega={4\Omega}/(2p+1)$.} 
	\label{3d_plot_det}
\end{figure*}
		
Next, we study an effect of detuning $\Delta\omega =  \omega_d - \omega_{01}$ varying the central frequency $\omega_d$ with fixed $\delta\omega$. We measure the spectral components of the signal at frequencies $\omega_{\pm (2p+1)}$ as a function of $\Omega_\pm$ in the condition of $\Omega_- = \Omega_+ = \Omega$. The observed effect is similar to Autler-Townes splitting, see Fig.~\ref{3d_plot_det} \cite{ATS_3LS}. 
The mixed signals of each order are split into two peaks with maxima at $\Delta\omega_{max}\approx\pm \zeta_p\Omega$, where $\zeta_p$ is a constant inversely proportional to $2p+1$. The peak position at a strong drive ($\Delta\omega \gg\Gamma_2$) can be estimated in a similar way as we have done it above for the peak maximum. To estimate $\Delta\omega_{max}$ we substitute the characteristic time $\Gamma_2^{-1}$ by $\tau \approx \Delta\omega^{-1}$. The peak maximum is then expected for $\zeta_p = \Delta\omega_{max}/\Omega \approx 4/(2p+1)$. 


{\C{Analysis.}} To quantitatively analyse the amplitudes of the wave mixing side peaks we consider the Hamiltonian of a single two-level system driven by two classical (coherent) waves 
\begin{multline}	
H = -\frac{\hbar \omega_{01}}{2}\sigma_z -\hbar \Omega_- \sigma_x \cos(\omega_d t - \delta\omega t)\\ -\hbar \Omega_+ \sigma_x \cos(\omega_d t + \delta\omega t),	
\end{multline} 
where $\Omega_+$ and $\Omega_-$ are amplitudes of the drives. First, we calculate a stationary solution for the master equation in the Rotating Wave Approximation. The term  $\delta\omega t$ is interpreted here as a slowly varying phase because $\delta\omega t \ll 1$ on a time scale $t \sim \Gamma_2^{-1}$. An analytical solution for the expectation value of the atomic annihilation operator is  
\begin{equation}
\langle \sigma^-\rangle = -\frac{\sin\theta}{\Lambda} \frac{\Omega_- e^{-i\delta\omega t} + \Omega_+ e^{i \delta\omega t}}{1 + \sin\theta\cos{2\delta\omega t}}. 
\label{exp_value}
\end{equation}
Here we introduce the following notations: $\theta = \arcsin\Big(\frac{2\Gamma_2 \Omega_- \Omega_+}{\Gamma_1 |\lambda|^2 + \Gamma_2(\Omega_-^2 + \Omega_+^2)}\Big)$, $\Lambda^{-1} = \frac{\lambda\Gamma_1}{4\Gamma_2 \Omega_-\Omega_+}$.
The denominator of Eq.~(\ref{exp_value}) can be rearranged according to  
\begin{equation}
\frac{1}{1+\frac{1}{2}\sin\theta(z + z^{-1})} = \frac{1}{\cos{\theta}}\Big(\frac{1}{1- yz} + \frac{1}{1- yz^{-1}}-1\Big), 
\label{exp_value2}
\end{equation}
where 
$z = e^{2i\delta\omega t}$ and $y = -\tan{\frac{\theta}{2}}$. We expand the right part of Eq. (\ref{exp_value2}) into power series in $z$ and arrive at 
\begin{equation}
\langle\sigma^-\rangle = -\frac{\Omega_- e^{-i\delta\omega t} + \Omega_+ e^{i \delta\omega t}}{\Lambda}\tan\theta \sum_{p=-\infty}^{\infty}y^{|p|} e^{i 2 p \delta\omega t}.
\label{exp_value3}
\end{equation}
Taking into account Eq.~(\ref{Vsc}) and transforming the sum to non-negative $p$, we obtain 
\begin{multline}
V^{sc} =  
-\frac{\hbar\Gamma_1\tan\theta}{\mu\Lambda}\sum_{p=0}^{\infty}y^p\Big[ (\Omega_- + y \Omega_+ ) e^{-i(2p+1)\delta\omega t}\\
+ (y \Omega_- + \Omega_+) e^{i(2p+1)\delta\omega t}\Big].
\label{an_eq}
\end{multline}
With the relations between the driving amplitude and the voltage amplitude $V_\pm \mu = \hbar\Omega_\pm$, we arrive to the analytical expression for the amplitude of each side spectral component 
\begin{equation}
V_{\pm(2p+1)}^{sc} =\frac{(-1)^p\Gamma_1\tan\theta\tan^p\frac{\theta}{2}}{\Lambda} (V_{\mp}\tan\frac{\theta}{2} - V_{\pm}),
\label{intensities}
\end{equation} 
which can be verified experimentally. Eq.~({\ref{intensities}}) shows that the factor $\tan^p\frac{\theta}{2}$ contains all the dependence of spectral components on the order $p$. To exemplify that, we also deduce ratios of consequent components of order $2p+1$ and $2p+3$ for the data presented in Fig.~\ref{Peaks_WM_with_fit} and present the result in the inset of Fig.~\ref{Peaks_WM_with_fit}(a) for $p=1,2,3$. Notice that it is the same for each pair independently of $p$, and fits well with the black solid line derived from Eq.(\ref{intensities}). The result is valid for classical coherent states, in which photon statistics is given by the Poissonian distribution.

{\C{Discussion.}}
We now compare the experimental data with our analytical expression of Eq.~(\ref{intensities}). Solid lines in Fig.~\ref{Peaks_WM_with_fit} show the calculated peak dependences, which are in a good agreement with the experimental data. Analysing Eq.~(\ref{intensities}) for extremums, we find that the peak maxima are well explained by asymptotic relation $\Omega_{max} \approx \sqrt{2} \Gamma_1(2p+1)/4$, which is consistent with our preliminary qualitative prediction and the physical picture we provide. Also, the driving amplitude dependence versus detuning well reproduces the measurement as it is shown in Fig.~\ref{3d_plot_det}. 
Quantitative analysis of Eq.~(\ref{intensities}) gives maximal response at 
$\omega_{max}/\Omega \approx 4/(2p+1)$.

Next, we illustrate that the wave mixing spectral components reveal photon statistics of the incident waves. In the strong coupling and weak driving regime ($\Omega_\pm \ll \Gamma_1$), the scattered photon number into the mode $\omega_{2p+1}$  in one direction is simplified from Eq.~(\ref{intensities}) to 
\begin{equation}
\langle N_{2p+1}\rangle \approx \langle N_{-} \rangle^{p} \langle N_{+} \rangle^{p+1},
\end{equation}
where 
$ \langle N_k \rangle = \Omega_k^2/\Gamma_1\Gamma_2$
is the mean photon number in the mode $\omega_k$ on the characteristic time interval $\tau= \Gamma_2^{-1}$. Remarkably, this is equivalent to the expectation value of the operator $(a_+ a^\dag_-)^p a_+$ averaged over the states $|\alpha_-\alpha_+\rangle$. Its squared value is $|\langle (a_+ a^\dag_-)^p a_+ \rangle|^2 \approx |(\alpha_-^{\ast})^p \alpha_+^{p+1}|^2 = \langle N_-\rangle^{p} \langle N_+\rangle^{p+1}$. The prefactors $(\alpha_-^{\ast})^p$ and $\alpha_+^{p+1}$ are determined by probability amplitudes of the corresponding photon states ($|p\rangle_-$ and $|p+1\rangle_+$). For instance, in the week driving regime $\alpha \ll 1$ and $\langle a^n \rangle \approx \bra{\alpha} a^{n} \frac{\alpha^{n}}{\sqrt{n!}} \ket{n} = \langle 0| \alpha^{n} |0\rangle = \alpha^{n}$, that is approximately equal to the probability amplitude of the photon-number state $\ket{n}$ in the coherent state $\ket{\alpha}$ multiplied by $\sqrt{n!}$.
The case with deviation from the classical coherent states has been already discussed though in a different regime of pulsed dynamics \cite{dmitriev2017quantum}. We suppose that our method is promising for detection and characterisation of non-classical coherent states, when the photon statistics deviates from the Poissonian one. 



{\C{Conclusions.}}
In conclusion, we have demonstrated a fundamental effect of wave mixing of stationary coherent states on a single two-level scatterer strongly coupled to a one-dimensional transmission line.      
We derive an analytical expression for the amplitudes of mixed states and have shown a series of other physical effects, for example, an Autler-Townes-like splitting of side peaks dependent on the number of scattered photons. The side peaks are results of multi-photon scattering processes and their amplitudes determined by the photon distribution in the coherent states. An interesting future application would be to visualize statistics of nonclassical coherent states. 
\begin{acknowledgements}
	This research is supported by the Russan Science Foundation, grant No. 16-12-00070. We are grateful to A. Semenov for useful discussions.
\end{acknowledgements}
	\bibliography{Bibl_lib}
\end{document}